\documentclass[doublecol,linenumbers]{epl2} 
\usepackage{amsmath,amsthm,amssymb,amscd}

\title{Work needed to drive a thermodynamic system between two distributions}
\shorttitle{Work needed to drive a thermodynamic system} 

\author{Yunxin Zhang\inst{1} }
\shortauthor{Y. Zhang}

\institute{
  \inst{1} Shanghai Key Laboratory for Contemporary Applied Mathematics, School of Mathematical Sciences, Fudan University, Shanghai 200433, China.
}

\abstract{
In this study, the minimum amount of work needed to drive a thermodynamic system from one initial distribution to another in a given time duration is discussed. Equivalently, for given amount of work, the minimum time duration required to complete such a transition is obtained. Results show that the minimum amount of work is used to achieve the following three objectives, to increase the internal energy of the system, to decrease the system entropy, to change the mean position of the system, and with other nonzero part dissipated into environment. To illustrate the results, an example with explicit solutions is presented.}

\begin{document}

\maketitle

\section{Introduction}
An essential difference between equilibrium and nonequilibrium systems is that the latter usually changes with time or requires extra energy to keep it in a steady state. Nonequilibrium systems are crucial in nature. For example, the human body is always maintained in a nonequilibrium steady state with the aid of ubiquitous molecular machines, including motor proteins (kinesin, dynein, and myosin) \cite{Howard2001,Vale2003}, DNA and RNA polymerase and ribosomes \cite{Cooper2000}, with energy stored in adenosine triphosphate, or with the difference in the chemical potential of ions \cite{Bray2001,Alberts2002,Sperry2007}. Macroscopically, motor vehicles can run steadily with the power provided by engines. While in mesoscopic scale, most of nonequilibrium systems are usually driven by molecular machines, which can be regarded as stochastic heat engines. Heat engine can work cyclically and extract work from the difference of temperature between two heat baths \cite{Sekimoto2010,Seifert2012Stochastic,Mart2016Colloidal,Giuliano2017}, where part of the heat $Q_h$ extracted from the hot heat bath with temperature $T_h$ is translated into work $W$, and the rest $Q_c=Q_h-W$ flows into the cold heat bath with temperature $T_c<T_h$.

According to the second law of thermodynamics, $Q_c/T_c-Q_h/T_h\ge0$, which indicates that the thermodynamic efficiency $\eta=W/Q_h=1-Q_c/Q_h\le\eta_C$, with $\eta_C:=1-T_c/T_h$ being the Carnot efficiency obtained firstly by Carnot two centuries ago \cite{Carnot1824}. Carnot efficiency $\eta_C$ can only be attained through a quasi static (reversible) process, with work duration $t$ being infinite and power $W/t$ vanishing. Efficiency $\eta<\eta_C$ when work duration $t$ is finite due to nontrivial energy dissipation. A previous study presented a method to optimise the performance of heat engines, including their power and efficiency, by reducing energy dissipation \cite{Zhang20191}. Using the same idea, this study discusses the inverse problem, that is, for a given work duration $t$, we determine how much work is required to drive a thermodynamic system from one distribution (state) to another and identity the optimal protocols.

This problem is crucial in nonequilibrium thermodynamics and was addressed recently in \cite{Schmiedl2007,Gomez2008,Aurell2011,Aurell2012,Sivak2012,Bonanca2014,Horowitz2018,Shiraishi2018}. In \cite{Schmiedl2007, Gomez2008}, for examples of over-damped and under-damped  Langevin dynamics, exact optimal protocols for perturbing the position and spring constant are derived. In \cite{Aurell2011,Aurell2012}, lower bound of dissipation is obtained generally within the theoretical framework of Langevin stochastic processes and using the method of Monge-Kantorovich optimal mass transport.
In \cite{Sivak2012}, the metric structure controlling the dissipation of finite time transformations within the linear response regime was discussed generally. In \cite{Bonanca2014}, it is found that energy dissipation can be written as a functional that depends only on the correlation time and fluctuations of the generalised force. In \cite{Horowitz2018}, finite time protocols that optimise the compromise between the standard deviation and mean of dissipated work were numerically determined for two canonical examples of driven mesoscopic systems. In \cite{Shiraishi2018}, a trade-off inequality between the speed of the state transformation and the entropy production was obtained.

Generally, during a driving process, external work is usually employed to perform the following: (1) increase the internal energy of the system, (2) decrease the system entropy (or equivalently, change the landscape of system distribution), (3) translate the mean center of the system, and (4) dissipate into environment. The work dissipation will always be nozero as long as there are entropy difference or mean center difference during the driving process, and the driving is not a quasi-static process. The main aim of this study is to find the optimal driving protocols, with which the work dissipation reaches it minimum.

\section{Optimal driving between two distributions}
According to the first law of thermodynamics, $\dot{E}=\dot{Q}+\dot{W}$, where $E$ is the internal energy of the system, $Q$ is the heat picked up by the system and $W$ is the work done to the system. Let $\rho(x,\tau)$ be the probability density to find the system at position (state) $x$ at time $\tau$, and $V(x,\tau)$ be the time-dependent external potential. For simplicity, variable $x$ is assumed to lie in the interval $[0,L]$. Under the assumption that the dynamics is Markovian, the internal energy $E$, heat flow $\dot{Q}$, input power $\dot{W}$, and system entropy $S(\tau)$ can be written as follows,
\begin{equation}\label{eq1}
\begin{aligned}
E(\tau)&=\int_0^{L}V(x,\tau)\rho(x,\tau)dx, \quad \cr \dot{Q}(\tau)&=\int_0^{L}\dot{\rho}(x,\tau)V(x,\tau)dx, \cr
\dot{W}(\tau)&=\int_0^{L}\rho(x,\tau)\dot{V}(x,\tau)dx,\quad\cr
S(\tau)&=-k_B\int_0^{L}\rho(x,\tau)\ln\rho(x,\tau)dx,
\end{aligned}
\end{equation}
where $k_B$ is the Boltzmann constant.
The time evolution of $\rho(x,\tau)$ satisfies the following Fokker-Planck equation \cite{Schmiedl2008Efficiency1,Lucia2015,Zhang20191},
\begin{eqnarray}\label{eq2}
\dot{\rho}(x,\tau)=-j'(x,\tau)=\left(\mu \rho(x,\tau)V'(x,\tau)+D\rho'(x,\tau)\right)'.
\end{eqnarray}
Here, $D$ is the diffusion constant; $\mu$ is the motility that satisfies $k_BT\mu=D$, with $T$ being the absolute temperature; and $j$ is the flux of probability density, which can be written as $j(x,\tau)=\rho(x,\tau)v(x,\tau)$, with $v(x,\tau)$ being the instantaneous velocity \cite{Zhang20091}. In this study, the dots indicate time derivatives, and the primes indicate derivatives according to variable $x$.

During time $0\le\tau\le t$, the heat $Q(t)$ flowing into the system is (see Eqs.~(\ref{eq1}, \ref{eq2})) {\small
\begin{equation}\label{eq2-1}
\begin{aligned}
Q(t)&=\int_0^t\dot{Q}(\tau)d\tau
=\int_0^t\int_0^{L}\dot{\rho}(x,\tau)V(x,\tau)dxd\tau\cr
&=-\int_0^t\int_0^{L}j'(x,\tau)V(x,\tau)dxd\tau\cr
&=\int_0^t\int_0^{L}j(x,\tau)V'(x,\tau)dxd\tau\cr
&=-\int_0^t\int_0^{L}\xi j(x,\tau)[v(x,\tau)+D(\ln \rho(x,\tau))'] dxd\tau\cr
=&-k_BT\int_0^t\int_0^{L}j(x,\tau)(\ln \rho(x,\tau))'dxd\tau\cr
&-\int_0^t\int_0^{L}\xi \rho(x,\tau)v^2(x,\tau)dxd\tau\cr
=&:T\Delta S(t)-W_{\rm diss}(t),
\end{aligned}
\end{equation}}
where $\xi=1/\mu$ is the drag coefficient, $\Delta S(t)=S(t)-S(0)$ is the change in entropy, and dissipation $W_{\rm diss}(t)$ is given as follows \cite{Schmiedl2008Efficiency1,Zhang20191,Zhang2019},
\begin{equation}\label{eq3}
\begin{aligned}
W_{\rm diss}(t)&=\int_0^t\int_0^{L}\xi\rho(x,\tau)v^2(x,\tau)dxd\tau\cr
&=\int_0^t\int_0^{L}\xi\dot{f}^2(x,\tau)/f'(x,\tau)dxd\tau,
\end{aligned}
\end{equation}
with $f(x,\tau)=\int_0^x\rho(z,\tau)dz$ being the distribution function. Note that, by the definition of total entropy production $\Delta S_{\rm tot}(t)=-Q(t)/T+\Delta S(t)$, we can easily find from Eq.~(\ref{eq2-1}) that $W_{\rm diss}(t)=T\Delta S_{\rm tot}(t)$. So $W_{\rm diss}(t)$ is actually the work dissipated into environment to increase entropy of the system \cite{Verley2014The,Holubec2015,Zhang20191}.

According to Eq.~(\ref{eq2}), distribution function $f(x,\tau)$ satisfies
\begin{eqnarray}\label{eq4}
\dot{f}(x,\tau)+v(x,\tau)f'(x,\tau)=0.
\end{eqnarray}

The same as in \cite{Zhang20191}, variation of $W_{\rm diss}(t)$ according to distribution function $f(x,t)$ is as follows,
\begin{small}
\begin{eqnarray}\label{eq4-1}
\delta W_{\rm diss}=\int_0^{t}\int_0^{L}2\xi\left[\frac{\dot{f}}{f'}
\partial_{x}\left(\frac{\dot{f}}{f'}\right)
-\partial_{\tau}\left(\frac{\dot{f}}{f'}\right)\right]
\delta f\,dx d\tau,
\end{eqnarray}
\end{small}
where $\delta f$ is an arbitrary variation of $f(x,\tau)$, which satisfies $\delta f(0, \tau)=\delta f(L,\tau)=\delta f(x,0)=\delta f(x,t)=0$. Since $\delta W_{\rm diss}=0$ at the minimum of $W_{\rm diss}$ for any variation $\delta f$, Eq.~(\ref{eq4-1}) indicates
$(\dot{f}/f')\partial_{x}(\dot{f}/f')-\partial_{\tau}(\dot{f}/f')=0$ for the optimal distribution function $f(x,\tau)=f^*(x,\tau)$. Due to Eq.~(\ref{eq4}), $\dot{f}/f'=-v$. Therefore, $\partial_{\tau}v(x,\tau)+v(x,\tau)\partial_{x}v(x,\tau)=0$. This means that, for the optimal distribution function $f^*(x,\tau)$, the slope $v^*(x,\tau)$ of its characteristic curves satisfies $dv^*(x,\tau)/d\tau\equiv0$. Therefore, the characteristic curves of Eq.~(\ref{eq4}) are all straight lines when dissipation $W_{\rm diss}(t)$ reaches its lower bound $W_{\rm diss}^*(t)$. Note, this result can also be obtained by the method presented in \cite{Aurell2011,Aurell2012}. This finding indicates that, lower bound $W_{\rm diss}^*(t)$ is attained when the characteristic curve of Eq.~(\ref{eq4}), which starts from $z$ at time $\tau=0$, is
\begin{eqnarray}\label{eq5}
x(z,\tau)=z+(\Gamma(z)-z)\tau/t,
\end{eqnarray}
where $\Gamma(z)$ is a map from interval $[0,L]$ to $[0,L]$ and satisfies $f(\Gamma(z),t)=f(z,0)$. The definition of characteristic curve means that, along any characteristic curves, $df(x(z,\tau),\tau)/d\tau=0$, therefore $f(x(z,\tau),\tau)=f(z,0)$ for any $0\le z\le L$ and $0\le\tau\le t$, for details see \cite{Zhang20191}.

For convenience, let $\rho_0(x):=\rho(x,0)$, $\rho_1(x):=\rho(x,t)$, $f_0(x):=f(x,0)$, $f_1(x):=f(x,t)$. For given probability densities $\rho_0(x)$ and $\rho_1(x)$, when dissipation $W_{\rm diss}(t)$ attains its lower bound $W_{\rm diss}^*(t)$, the characteristic curve of distribution function $f(x,\tau)$ is given by Eq.~(\ref{eq5}), where $\Gamma(z)$ is determined by $\rho_0(x)$ and $\rho_1(x)$ (or by $f_0(x)$ and $f_1(x)$ equivalently). Thus, the distribution function $f(x,\tau)$ for any time $0\le\tau\le t$ and $0\le x\le L$ can be obtained. In fact, $f(x,\tau)=f_0(z)$, with $z$ determined by $x(z,\tau)=x$. Consequently, probability density $\rho(x,\tau)$ can be obtained by $\rho(x,\tau)=f'(x,\tau)$. The optimal potential $V^*(x,\tau)$, which determines the thermodynamic process, can be derived with Eq.~(\ref{eq2}), for detailed formulations see \cite{Zhang20191}.

This study uses potential $V(x,\tau)$ as a protocol to reduce dissipation $W_{\rm diss}(t)$ and make it as low as possible. For detailed analysis and concrete examples, see \cite{Zhang20191,Zhang2019}. In brief, for any given probability densities $\rho_0(x)$ and $\rho_1(x)$, and duration $t$, we can select one specific potential $V^*(x,\tau)$, with which dissipation $W_{\rm diss}(t)$ reaches its lower bound $W_{\rm diss}^*(t)$. From Eqs.~(\ref{eq3}, \ref{eq5}) and the definition of the characteristic curve, the lower bound of dissipation $W_{\rm diss}^*(t)$ can be obtained as follows \cite{Zhang20191}, {\small
\begin{equation}\label{eq6}
\begin{aligned}
W_{\rm diss}^*(t)&=\int_0^t\int_0^{L}\xi\rho(x,\tau)v^2(x,\tau)dxd\tau\cr
&=\int_0^t\int_0^{L}\xi\rho(x(z,\tau),\tau)v^2(x(z,\tau),\tau)x'(z,\tau)dzd\tau\cr
&=\int_0^t\int_0^{L}\xi f'(x(z,\tau),\tau)x'(z,\tau)\left(\frac{\Gamma(z)-z}{t}\right)^2dzd\tau\cr
&=\int_0^t\int_0^{L}\xi f'(z,0)\left(\frac{\Gamma(z)-z}{t}\right)^2dzd\tau\cr
&=\int_0^t\int_0^{L}\xi\rho_0(z)\left(\frac{\Gamma(z)-z}{t}\right)^2dzd\tau\cr
&=\frac{\xi}{t}\int_0^{L}\rho_0(z)(\Gamma(z)-z)^2dz.
\end{aligned}
\end{equation}
According to the first law of thermodynamics, the minimum amount of work required to drive the system from probability density $\rho_0(x)$ to $\rho_1(x)$ in duration $t$ is
\begin{eqnarray}\label{eq7}
W^*(t)=\Delta E-Q(t)=\Delta E-T\Delta S+W_{\rm diss}^*(t),
\end{eqnarray}
where $\Delta E$ is the change in internal energy, $\Delta S$ is the increase in entropy and $W_{\rm diss}^*(t)$ is the lower bound of dissipation (total entropy production \cite{Zhang20191}). Eqs.~(\ref{eq6}, \ref{eq7}) show that for given work input $W$, the minimum of time duration $t$ needed to drive the system from density $\rho_0(x)$ to $\rho_1(x)$ is
\begin{eqnarray}\label{eq7-1}
t^*(W)=\frac{\xi \int_0^{L}\rho_0(z)(\Gamma(z)-z)^2dz}{W-\Delta E+T\Delta S}.
\end{eqnarray}
Obviously, $t^*(W)$ decreases with work input $W$ and change in entropy $\Delta S$ but increases with $\Delta E$ and friction $\xi$. Notably, $\Delta E$ and $\Delta S$ may not be positive. Therefore, work input $W$ may be negative, which is the case of heat engines.

Generally, the work $W$ required to complete the transition of a thermodynamic system from density $\rho_0(x)$ to density $\rho_1(x)$ in duration $t$ is not less than $W^*(t)$, $W\ge W^*(t)$. Combined with Eqs.~(\ref{eq6}, \ref{eq7}), this implies that,
\begin{eqnarray}\label{eq7-2}
W-\Delta E+T\Delta S\ge\xi \left.\left(\int_0^{L}\rho_0(z)(\Gamma(z)-z)^2dz\right)\right/t.
\end{eqnarray}
This can be regarded as an {\it uncertainty principle} as in quantum theory \cite{Deffner2017}.

It may need to point out that the lower bound of dissipation $W_{\rm diss}^*(t)$ changes with $t$ like $1/t$ is because that, in our theoretical framework, the friction is assumed to be proportional to velocity. Generally, if the friction is $\xi v^{\sigma}$, then the lower bound of dissipation, denoted by $W_{\rm diss}^*(t;
\sigma)$, will change with $t$ like $1/t^{\sigma}$. Actually, by the same process as above, it can be shown that $W_{\rm diss}^*(t; \sigma)=\xi\left.\left(\int_0^{L}\rho_0(z)(\Gamma(z)-z)^{\sigma+1}dz\right)\right/t^{\sigma}$.

\section{Causes of dissipation}
To show detailed compositions of dissipation $W_{\rm diss}^*(t)$, we denote
\begin{eqnarray}\label{eq7-3}
c_{0}:=\int_0^{L}z\rho_{0}(z)dz,\quad
c_{1}:=\int_0^{L}z\rho_{1}(z)dz,
\end{eqnarray}
and $\Delta c:=c_1-c_0$.
Here, $c_0$ and $c_1$ are mean centers of the system at time $\tau=0$ and $\tau=t$, respectively, and $\Delta c$ is the change in mean system position during the whole driven process. By definition, $f_0(z)=f_1(\Gamma(z))$. It indicates that $\rho_0(z)=\rho_1(\Gamma(z))\Gamma'(z)$ and $\rho_1(z)=\rho_0(\Gamma^{-1}(z))/\Gamma'(\Gamma^{-1}(z))$. Thus,
\begin{eqnarray}\label{eq7-4}
\begin{aligned}
c_1=&\int_0^{L}z\rho_1(z)dz=\int_0^{L}z\rho_0(\Gamma^{-1}(z))/\Gamma'(\Gamma^{-1}(z))dz\cr
=&\int_0^{L}\Gamma(z)\rho_0(z)dz,
\end{aligned}
\end{eqnarray}
where the last equality is obtained by variable change $z=\Gamma(y)$.
By Eqs.~(\ref{eq7-3}, \ref{eq7-4}),
$$\Delta c=c_1-c_0=\int_0^{L}(\Gamma(z)-z)\rho_0(z)dz.$$

From the definition of characteristic curve $x(z,\tau)$, which satisfies $f_0(z)=f(x(z,\tau),\tau)$ for any $\tau\ge0$, we obtain $\rho_0(z)=\rho(x(z,\tau),\tau)x'(z,\tau)$ and $\rho(z,\tau)=\rho_0(x^{-1}(z,\tau))/x'(x^{-1}(z,\tau), \tau)$.
The instantaneous velocity $v(z,\tau)$ at position $z$ and at time $\tau$ equals to the slope of the characteristic curve through point $(z, \tau)$, from Eq.~(\ref{eq5}), $v(z,\tau)=[\Gamma(x^{-1}(z,\tau))-x^{-1}(z,\tau)]/t$.
Therefore, the mean flux of probability at time $\tau$ is {\small
\begin{eqnarray}\label{eq7-5}
\begin{aligned}
J(\tau):=&\int_0^{L}j(z,\tau)dz=\int_0^{L}\rho(z,\tau)v(z,\tau)dz\cr
=&\int_0^{L}\frac{\rho_0(x^{-1}(z,\tau))}{x'(x^{-1}(z,\tau), \tau)}
\frac{\Gamma(x^{-1}(z,\tau))-x^{-1}(z,\tau)}{t}dz\cr
=&\frac1t\int_0^{L}(\Gamma(y)-y)\rho_0(y)dy=\frac{\Delta c}{t}.
\end{aligned}
\end{eqnarray}}
Eq.~(\ref{eq7-5}) indicates that with the optimal potential $V^*(x,\tau)$, probability flux $J(\tau)\equiv J:=\Delta c/t$ is constant, and $\Delta c=\int_0^tJ(\tau)d\tau=Jt$. According to Schwartz inequality,
$$
\begin{aligned}
(Jt)^2=&(\Delta c)^2=\left[\int_0^{L}\rho_0^{1/2}(z)\rho_0^{1/2}(z)(\Gamma(z)-z)dz\right]^2dz\cr
\le&\int_0^{L}\rho_0(z)(\Gamma(z)-z)^2dz
=\frac{tW_{\rm diss}^*(t)}{\xi},
\end{aligned}
$$
which means $\xi J^2t=\xi (\Delta c)^2/t\le W_{\rm diss}^*(t)$, or equivalently
$$\frac{DW_{\rm diss}^*(t)}{J^2t}\ge k_BT.$$
This is a special case of thermodynamic uncertainty relations, as obtained in \cite{Barato2015,Gingrich2016,Pietzonka2017,Macieszczak2018}. Intuitively, $\xi (\Delta c)^2/t=\xi J^2t=(\xi Jt)J=(\xi \bar{V})J$ is the {\it mechanical} part of $W_{\rm diss}^*(t)$, which is needed to drive the system from mean position $c_0$ to mean position $c_1$. Here $\bar{V}:=\int_0^t\int_0^{L}\rho(z,\tau)v(z,\tau)dzd\tau=\int_0^t J(\tau) d\tau=Jt$ is the mean translocation velocity of the whole driving process. In the following, we denote
\begin{eqnarray}\label{eq7-6}
W_{\rm mech}^*(t):= \xi J^2t=\frac{\xi (\Delta c)^2}{t}.
\end{eqnarray}

From discussions above, $\int_0^{L}(z-c_0)\rho_0(z)dz=\int_0^{L}(\Gamma(z)-c_1)\rho_0(z)dz=0$. Hence, dissipation $W_{\rm diss}^*(t)$ can be decomposed into the following two parts (see Eq.~(\ref{eq6})),
\begin{eqnarray}\label{eq8}
W_{\rm diss}^*(t)=W_{\rm therm}^*(t)+W_{\rm mech}^*(t),
\end{eqnarray}
where the {\it thermal} part is given by
\begin{equation}\label{eq9}
\begin{aligned}
W_{\rm therm}^*(t)&=\frac{\xi}{t}\int_0^{L}\rho_0(z)[\Gamma(z)-z-\Delta c]^2dz\cr
&=\frac{\xi}{t}\int_0^{L}\rho_0(z)[(\Gamma(z)-c_1)-(z-c_0)]^2dz\cr
&=\frac{\xi}{t}(\delta_0^2+\delta_1^2-2\delta_{\rm cov}),
\end{aligned}
\end{equation}
with
\begin{equation}\label{eq10}
\begin{aligned}
\delta_0:=&\left(\int_0^{L}\rho_0(z)(z-c_0)^2dz\right)^{1/2},\cr
\delta_1:=&\left(\int_0^{L}\rho_0(z)(\Gamma(z)-c_1)^2dz\right)^{1/2}\cr
=&\left(\int_0^{L}\rho_1(z)(z-c_1)^2dz\right)^{1/2},\cr
\delta_{\rm cov}:=&\int_0^{L}\rho_0(z)(z-c_0)(\Gamma(z)-c_1)dz\cr
=&\int_0^{L}\rho_1(z)(\Gamma^{-1}(z)-c_0)(z-c_1)dz.
\end{aligned}
\end{equation}

According to Schwartz inequality, $\delta_{\rm cov}\le\delta_0\delta_1$. So $W_{\rm therm}^*(t)\ge(\delta_1-\delta_0)^2\xi/t\ge0$. Meanwhile, $W_{\rm therm}^*(t)=0$ only if $\Gamma(z)=z+\Delta c=z+c_1-c_0$, which is equivalent to $\rho_0(z)=f'_0(z)=f'_1(\Gamma(z))\Gamma'(x)=\rho_1(\Gamma(z))=\rho_1(z+\Delta c)$. Hence, the {\it thermal} part of dissipation $W_{\rm therm}^*(t)$ vanishes only when the final density $\rho_1(z)$ of the system is just the translation of the initial density $\rho_0(z)$. For such special cases, $W_{\rm diss}^*(t)=W_{\rm mech}^*(t)=\xi (\Delta c)^2/t$, dissipation is only due to the mechanical translation of the system. Evidently, $W_{\rm mech}^*(t)=0$ only when there is no mechanical translation $\Delta c=c_1-c_0=0$, or no friction $\xi=0$, or duration $t\to\infty$. The {\it thermal} part of dissipation $W_{\rm therm}^*(t)$ is the minimum amount of energy used to change the landscape of system distribution.

The {\it thermal} part of dissipation can be rewritten as follows,
\begin{eqnarray}\label{eq11}
W_{\rm therm}^*(t)=W_{\rm therm}^0(t)-2\delta_{\rm cov}\xi/t,
\end{eqnarray}
where we denote $W_{\rm therm}^0(t):=(\delta_0^2+\delta_1^2)\xi/t$. Thus, from Eq.~(\ref{eq8}),
\begin{equation}\label{eq12}
\begin{aligned}
W_{\rm diss}^*(t)&=W_{\rm mech}^*(t)+W_{\rm therm}^0(t)-2\delta_{\rm cov}\xi/t\cr
&=:W_{\rm diss}^0(t)-2\delta_{\rm cov}\xi/t.
\end{aligned}
\end{equation}
It can be shown that $W_{\rm therm}^0(t)=(\delta_0^2+\delta_1^2){\xi}/{t}$
\begin{equation}\label{eq13}
\begin{aligned}
=&\frac{\xi}{t}\int_0^{L}\int_0^{L}\rho_0(z)\rho_1(w)[(w-c_1)-(z-c_0)]^2dzdw\cr
=&\frac{\xi}{t}\int_0^{L}\int_0^{L}\rho_0(z)\rho_1(w)(w-z-\Delta c)^2dzdw,
\end{aligned}
\end{equation}
and $W_{\rm diss}^0(t)=W_{\rm therm}^0(t)+W_{\rm mech}^*(t)$ {\small
\begin{equation}\label{eq14}
\begin{aligned}
=&\frac{\xi}{t}\left(\int_0^{L}\int_0^{L}\rho_0(z)\rho_1(w)(w-z-\Delta c)^2dzdw+(\Delta c)^2\right)\cr
=&\frac{\xi}{t}\int_0^{L}\int_0^{L}\rho_0(z)\rho_1(w)(w-z)^2dzdw\cr
=&\int_0^t\int_0^{L}\int_0^{L}\xi\rho_0(z)\rho_1(w)\left(\frac{w-z}{t}\right)^2dzdwd\tau.
\end{aligned}
\end{equation}}

The last term in Eq.~(\ref{eq14}) indicates that $W_{\rm diss}^0(t)$ is the minimum of work dissipation if the transition from density $\rho_0(x)$ to density $\rho_1(x)$ is completely random, where the system at any initial state (position) has the same probability density $\rho_1(w)$ to finally reach state (position) $w$, but with the optimal transition trajectory (see Eq.~(\ref{eq6})). By contrast, $W_{\rm diss}^*(t)$ is the minimum of work dissipation corresponding to the optimal cases, where the system at initial state (position) $x$ reaches the final state (position) $\Gamma(x)$  definitely, also with the optimal transition trajectory (see Eq.~(\ref{eq6})). The meaning of $W_{\rm therm}^0(t)$ given in Eq.~(\ref{eq13}) is similar to that of $W_{\rm diss}^0(t)$, but with the deduction of {\it mechanical} part $W_{\rm mech}^*(t)$.

Showing that $W_{\rm diss}^*(t)\le W_{\rm diss}^0(t)$ is equivalent to showing that $W_{\rm therm}^*(t)\le W_{\rm therm}^0(t)$. With Eq.~(\ref{eq11}), it is equivalent to showing that $\delta_{\rm cov}\ge0$.
By definition, $c_0=\int_0^{L}z\rho_0(z)dz$ and $c_1=\int_0^{L}z\rho_1(z)dz=
\int_0^{L}\Gamma(z)\rho_0(z)dz$. Hence, $\delta_{\rm cov}$ can be reformulated as $\delta_{\rm cov}=\int_0^{L}z\Gamma(z)\rho_0(z)dz-c_0c_1$ (see Eq.~(\ref{eq10})). Therefore,{\small
\begin{eqnarray}\label{eq15}
\delta_{\rm cov}&=&\int_0^{L}z\Gamma(z)\rho_0(z)dz-c_0c_1\cr
&=&\frac{1}{2}\left(\int_0^{L}\int_0^{L}z\Gamma(z)\rho_0(z)\rho_0(w)dwdz\right.\cr
&&+\int_0^{L}\int_0^{L}w\Gamma(w)\rho_0(w)\rho_0(z)dwdz\cr
&&-\int_0^{L}\int_0^{L}z\Gamma(w)\rho_0(z)\rho_0(w)dwdz\cr
&&\left.-\int_0^{L}\int_0^{L}w\Gamma(z)\rho_0(w)\rho_0(z)dwdz\right)\cr
&=&\frac{1}{2}\int_0^{L}\int_0^{L}(z-w)[\Gamma(z)-\Gamma(w)]\rho_0(z)\rho_0(w)dwdz\cr
&\ge&0,
\end{eqnarray}}
where the last inequality is due to the fact that $\Gamma(x)$ is an increasing function of $x$, $\Gamma'(x)=\rho_0(x)/\rho_1(\Gamma(x))\ge0$.

Finally, we remark that the definition of $\Gamma(x)$ is straight forward if both $\rho_0(x)$ and $\rho_1(x)$ are greater than zero for almost everywhere in the interval $[0, L]$. For general cases where either $\rho_0(x)$ or $\rho_1(x)$, or both of them are equal to zero in one or several subintervals of $[0,L]$, the definition of $\Gamma(x)$ is not unique; nevertheless, all the analyses and results given in this study still hold, and function $\Gamma(x)$ can always be constructed as an increasing function.

Although this study assumes that variable $x$ lies in the interval $[0, L]$, the same results can be obtained for any other types of domain, such as $[0,\infty)$, $(-\infty, \infty)$, $[a, b]$ for any $a<b$, or any other domains composed of multiple subintervals. Similar to \cite{Zhang20191}, the influence of interval scale $L$ can be analysed by rescaling probability $\rho_{0/1}(x)$ into interval $[0,1]$.

\begin{figure}[h]
	\centering
	\includegraphics[width=0.49\textwidth]{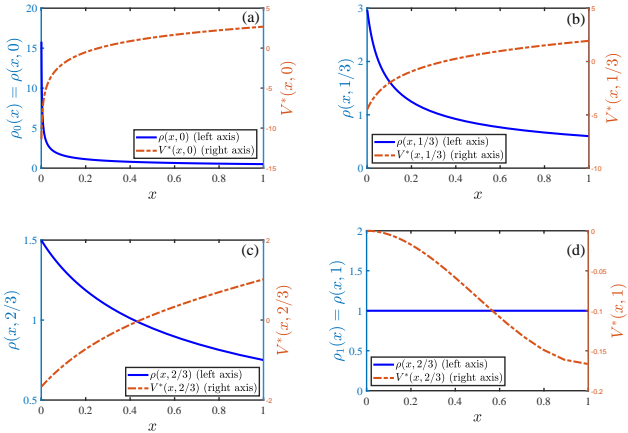}
	\caption{The optimal potential $V^*(x,\tau)$ (dashed lines with right axes) and probability density $\rho(x,\tau)$  (solid lines with left axes) for the illustrative example, with time $\tau=0, 1/3, 2/3, 1$, respectively. See Eqs.~(\ref{eq16}, \ref{eq17}) for explicit expressions of $\rho(x,\tau)$ and $V^*(x,\tau)$ (with $C(\tau)\equiv0$ in calculations). Values of parameters used in calculations are $L=1, t=1$ and $\xi=1, k_BT=1$.
}\label{Fig1}
\end{figure}
\section{Illustrative example}
To illustrate the results obtained above, we present an example with an explicit solution (see also \cite{Zhang20191} for more examples). Let $L=1$, driving period $t=1$, the initial probability density $\rho_0(x)=1/(2\sqrt{x})$, and the final probability density $\rho_1(x)=1$. Then from the relation $\rho_0(x)=\rho_1(\Gamma(x))\Gamma'(x)$, we can obtain the map $\Gamma(x)=\sqrt{x}$.
Below we will calculate work dissipations of the optimal driving process from probability density $\rho_0(x)$ to probability density $\rho_1(x)$ in time $t=1$.

From Eq.~(\ref{eq5}), the characteristic curve of Eq.~(\ref{eq4}), which starts from $z$ at time $\tau=0$, is
$$
x(z,\tau)=z+(\sqrt{z}-z)\tau,
$$
with which the probability $\rho(z,\tau)$ at any time $\tau$ can be obtained as follows,
\begin{equation}\label{eq16}
\rho(z,\tau)=\frac{\rho_0(x^{-1}(z,\tau))}{x'(x^{-1}(z,\tau),\tau)}=\frac{1}{\sqrt{\tau^2+4(1-\tau)z}}.
\end{equation}
For further details, see \cite{Zhang20191}. By Eq.~(\ref{eq2}), potential
\begin{equation}\label{eq16-1}
V(x,\tau)=\int_0^x\frac{\xi\int_0^z\partial_\tau \rho(y,\tau)dy-k_BT \partial_z \rho(z,\tau)}{\rho(z,\tau)}dz+C(\tau).
\end{equation}
It can be shown that, for the optimal driving process, potential is as follows,
{\small
\begin{equation}\label{eq17}
\begin{aligned}
V^*(x,\tau)=&\xi\left.\left[\frac12(1-\tau)y^2-\left(\frac23-\tau\right)y^{\frac32}
-\frac{\tau}{2}y\right]\right|_{y=\Gamma^{-1}_{\tau}(x)}\cr
&+k_BT\ln\sqrt{\tau^2+4(1-\tau)x}+C(\tau),
\end{aligned}
\end{equation}}
where $\Gamma^{-1}_{\tau}(x)=[(\sqrt{\tau^2+4(1-\tau)x}-\tau)/(2(1-\tau))]^2$ is the inverse of characteristic curve $x(z,\tau)$ given in Eq.~(\ref{eq5}). In this example, we assume $C(\tau)\equiv0$ for simplicity. See Fig.~\ref{Fig1} for plots of potential $V^*(x,\tau)$ and probability density $\rho(x,\tau)$ with time $\tau=0, 1/3, /2/3, 1$, respectively.

It can be easily shown that,  {\small
\begin{eqnarray}\label{eq18}
&&V^*(x,0)=\xi\left(\frac12x^2-\frac23x^{\frac32}\right)+k_BT\ln2\sqrt{x},\cr
&&V^*(x,1)=\xi\left(\frac13x^3-\frac12x^{2}\right).
\end{eqnarray}}
Therefore, {\small
\begin{eqnarray}\label{eq19}
&&E(0)=\int_0^1V^*(x,0)\rho_0(x)dx=(\ln2-1)k_BT-\frac{\xi}{15},\cr
&&E(1)=\int_0^1V^*(x,1)\rho_1(x)dx=-\frac{\xi}{12},\cr
&&S(0)=-k_B\int_0^1\rho_0(x)\ln \rho_0(x)dx
=(\ln2-1)k_B,\cr
&&S(1)=-k_B\int_0^1\rho_1(x)\ln \rho_1(x)dx=0.
\end{eqnarray}}
So, the change in internal energy is $\Delta E=E(1)-E(0)=(1-\ln2)k_BT-\xi/60$,
and the change in entropy is $\Delta S=S(1)-S(0)=(1-\ln2)k_B$.
The dissipation $W^*_{\rm diss}(t)$ for this example is
\begin{eqnarray}\label{eq20}
W^*_{\rm diss}(t=1)=\xi\int_0^{1}\rho_0(z)[\Gamma(z)-z]^2dz=\frac{\xi}{30}.
\end{eqnarray}

Mean center $c_0=\int_0^1z\rho_0(z)dz=1/3$, and $c_1=\int_0^1z\rho_1(z)dz=1/2$. Thus, $\Delta c=c_1-c_0=1/6$, and the {\it mechanical} part of dissipation $W^*_{\rm mech}(t=1)=\xi (\Delta c)^2/t=\xi/36$. The {\it thermal} part of dissipation $W^*_{\rm therm}(t=1)=W^*_{\rm diss}(t=1)-W^*_{\rm mech}(t=1)=\xi/180$.
Meanwhile, it can be easily shown that $\delta_0^2=4/45$, $\delta_1^2=1/12$, so $W^0_{\rm therm}(t=1)=\xi(\delta_0^2+\delta_1^2)/t=31\xi/180$. Obviously, $W^0_{\rm therm}(t=1)$ is much larger than the {\it thermal} part of dissipation $W^*_{\rm therm}(t=1)$.

Finally, the minimum work needed to drive the system from probability density $\rho_0(x)$ to probability density $\rho_1(x)$ in duration $t=1$ is (see Eq.~(\ref{eq7}))
\begin{eqnarray}\label{eq21}
W^*(t=1)=\Delta E-T\Delta S+W_{\rm diss}^*(t=1)=\frac{\xi}{60}.
\end{eqnarray}
\noindent So the minimum work required to drive the system from probability density $\rho_0(x)$ to probability density $\rho_1(x)$ in duration $t=1$ is $\xi/60$. It can be easily shown that the minimum work required to drive the system inversely, {\it i.e.}, from probability density $\rho_1(x)$ to probability density $\rho_0(x)$ in duration $t=1$, is $\xi/20$, in which the minimum dissipation $W_{\rm diss}^*(t=1)$ is the same, $W_{\rm diss}^*(t=1)=\xi\int_0^{1}\rho_1(z)[\Gamma^{-1}(z)-z]^2dz=\xi/30$, while the change in $E-TS$ is $\Delta E-T\Delta S=\xi/60$.

To show more details about this illustrative example, the change over time of system entropy $S(\tau)$, mean internal energy $E(\tau)$, mean center $c(\tau):=\int_0^{L}x\rho(x,\tau)dx$, and the cumulation value of work dissipation $W^*_{\rm diss}(\tau):=\int_0^{\tau}\int_0^{L}\xi\rho_0(z)\textbf{(}[x(z,\tau)-z]/\tau\textbf{)}^2dzd\tau$
are displayed in Fig.~\ref{Fig2}. Calculation results show that both $c(\tau)$ and $W^*_{\rm diss}(\tau)$ change linearly with time $\tau$. This indicates that, for the optimal driving process between two given distributions, the translocation velocity of mean system center and the rate of work dissipation (or the total entropy production rate) are always constants.

\begin{figure}[h]
	\centering
	\includegraphics[width=0.49\textwidth]{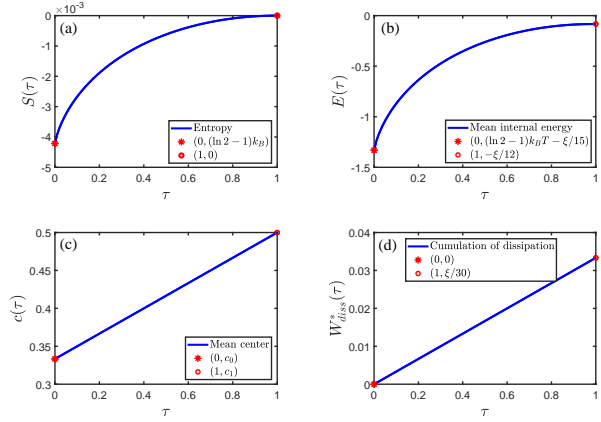}
	\caption{The change over time of system entropy $S(\tau):=-k_B\int_0^{L}\rho(x,\tau)\ln\rho(x,\tau)dx$, mean internal energy $E(\tau):=\int_0^{L}V(x,\tau)\rho(x,\tau)dx$, mean center $c(\tau):=\int_0^{L}x\rho(x,\tau)dx$, and the cumulation value of work dissipation $W^*_{\rm diss}(\tau):=\int_0^{\tau}\int_0^{L}\xi\rho_0(z)\textbf{(}[x(z,\tau)-z]/\tau\textbf{)}^2dzd\tau$
for the illustrative example. Note, from Eqs.~(\ref{eq5}, \ref{eq6}), $W^*_{\rm diss}(\tau)$ can be reformulated as $W^*_{\rm diss}(\tau)=\tau\xi\int_0^{L}\rho_0(z)[\Gamma(z)-z]^2dz/t^2=\tau W^*_{\rm diss}/t$. Meanwhile, from $\rho(z,\tau)=\rho_0(x^{-1}(z,\tau))/x'(x^{-1}(z,\tau), \tau)$, $c(\tau)$ can be reformulated as $c(\tau)=c_0+(c_1-c_0)\tau/t$ (this can be verified directly by the explicit expression of $\rho(x,\tau)$ given in Eq.~(\ref{eq16})).    }\label{Fig2}
\end{figure}

\section{Summary}
The lower bound of work needed to drive a thermodynamic system between two distributions is discussed in this study. During the driving process, part of the work is used (output) to increase (decrease) the internal energy, part is used (output) to decrease (increase) the system entropy, and the rest is dissipated during the change in landscape of system probability density and the translation of system mean position. Amongst these, dissipation can be optimised to its minimum, which is proportional to the drag coefficient and inversely proportional to the driving duration. Roughly speaking, the lower bound of dissipation depends on mean values and variances of the initial and final distributions of the system and on their covariance. For a given work input, the minimum time required to complete the driving process is also obtained. It decreases with the work input and change in system entropy but increases with the friction coefficient and change in system internal energy. As pointed out previously that similar results about the lower bound of dissipation have been obtained previously in \cite{Aurell2011,Aurell2012}. The main difference between the method used in this study and the one in \cite{Aurell2011,Aurell2012} is that, in this study, the lower bound is obtained by the method of characteristics and variation, and within the framework of Fokker-Planck equation. Moreover, in this study, causes of dissipation are also discussed in details.

\begin{acknowledgments}
The author thanks S. Deffner of UMBC for his comments and suggestions.
\end{acknowledgments}

\end{document}